\documentclass[letterpaper, 12pt]{article}
\usepackage{jheppub}

\usepackage{graphicx}
\usepackage{bbm}
\usepackage{float}
\usepackage{amsthm}
\usepackage{rotating}
\usepackage{amssymb}
\usepackage{physics}
\usepackage{upgreek}
\usepackage{amsmath}
\usepackage{mathtools}
\usepackage{multirow}
\usepackage{afterpage}
\usepackage{epstopdf}
\usepackage{mathrsfs}
\usepackage[labelformat=simple]{subcaption}
\usepackage{calligra}
\usepackage[T1]{fontenc}  

\begin{document}

\begin{titlepage}

\setcounter{page}{1} \baselineskip=15.5pt \thispagestyle{empty}
{\flushright {ITP-CAS-25-215}\\}
		
\bigskip\
		
\vspace{0.9cm}%%%%%%%%
\begin{center}
%{%\fontsize{20}{28}
%{\LARGE \bfseries the\vspace{0.24cm}\\ the}
{\LARGE \bfseries Completing Axion Double Level Crossings}
\end{center}
%\vspace{0.5cm}
\vspace{0.15cm}
			
\begin{center}
{\fontsize{14}{30}\selectfont Hai-Jun Li$^{a,b,c}$, Wei Chao$^{d,e}$, Huai-Ke Guo$^{c,f}$, and Yu-Feng Zhou$^{b,c,f,g}$}
\end{center}
%\vspace{0.1 cm}
\begin{center}
\vspace{0.25 cm}
\textsl{$^a$Department of Physics, Yunnan University, Kunming 650091, China}\\
\textsl{$^b$Institute of Theoretical Physics, Chinese Academy of Sciences, Beijing 100190, China}\\
\textsl{$^c$International Centre for Theoretical Physics Asia-Pacific, Beijing 100190, China}\\
\textsl{$^d$Center for Advanced Quantum Studies, School of Physics and Astronomy, Beijing Normal University, Beijing 100875, China}\\
\textsl{$^e$Key Laboratory of Multi-scale Spin Physics, Ministry of Education, Beijing Normal University, Beijing 100875, China}\\
\textsl{$^f$School of Physical Sciences, University of Chinese Academy of Sciences, Beijing 100049, China}\\
\textsl{$^g$School of Fundamental Physics and Mathematical Sciences, Hangzhou Institute for Advanced Study, UCAS, Hangzhou 310024, China}\\

\vspace{-0.1 cm}				
\begin{center}
{E-mail: \textcolor{blue}{\tt {lihaijun@itp.ac.cn}}, \textcolor{blue}{\tt {chaowei@bnu.edu.cn}}, \textcolor{blue}{\tt {guohuaike@ucas.ac.cn}}, \textcolor{blue}{\tt {yfzhou@itp.ac.cn}}}
\end{center}	
\end{center}
\vspace{0.6cm}
\noindent

In this work, we present the refinement of axion double level crossings within the context of multi-axion mass mixing, specifically focusing on cases where the number of axions exceeds two.
Our investigation reveals that double level crossings are a common phenomenon in the mass mixing of the $Z_{\mathcal N}$ axion and axion-like particles.
Physically, these double level crossings involve a first level crossing at high temperatures followed by a second level crossing induced by the $Z_{\mathcal N}$ axion mass transition at $T_{\rm QCD}$.
We introduce the general model for double level crossings, along with several toy examples, and redefine the light and heavy axion scenarios. 
In the light axion scenario, double level crossings can occur multiple times in the large ${\mathcal N}$ limit.
However, excessively large values of ${\mathcal N}$ may also prevent the occurrence of double level crossings.
Conversely, in the heavy axion scenario, excessively small ${\mathcal N}$ may similarly prevent their occurrence.
Our findings also have some intriguing implications for axion cosmology.
 		
\vspace{1.1cm}%%%%%%%%
			
\bigskip
\noindent\today
\end{titlepage}
			
\setcounter{tocdepth}{2}
			
%\hrule
%\tableofcontents			
%\bigskip\medskip
%\hrule
%\bigskip\bigskip
%\pagebreak\
 
%\newpage

\section{Introduction}%%%%%%%%%%%%%%%%%%%%%%%%%%Introduction

String theory predicts the existence of a plenitude of axions \cite{Svrcek:2006yi, Conlon:2006tq}, constituting what is known as the String Axiverse, where the QCD axion coexists with a large number of ultra-light axion-like particles (ALPs) \cite{Arvanitaki:2009fg, Acharya:2010zx, Cicoli:2012sz, Demirtas:2018akl, Reig:2021ipa, Demirtas:2021gsq}.
The corresponding low-energy effective Lagrangian is given by
\begin{eqnarray}
\mathcal{L}\supset\dfrac{1}{2}\sum_i f_{\theta_i}^2 \left(\partial \theta_i \right)^2-\sum_i\Lambda_i^4\left[1-\cos\left(\sum_j n_{ij} \theta_j \right)\right]\, ,
\end{eqnarray} 
where $\theta_i$, $f_{\theta_i}$, $\Lambda_i$, and $n_{ij}$ represent the axion angles, axion decay constants, overall scales, and domain wall numbers, respectively.
The QCD axion was initially predicted by the Peccei-Quinn (PQ) mechanism to dynamically solve the strong CP problem in the Standard Model (SM) \cite{Peccei:1977hh, Peccei:1977ur, Weinberg:1977ma, Wilczek:1977pj}.
It can also serve as a natural candidate for cold dark matter (DM), being non-thermally produced in the early Universe via the misalignment mechanism \cite{Preskill:1982cy, Abbott:1982af, Dine:1982ah}.
See $\rm e.g.$ refs.~\cite{Marsh:2015xka, DiLuzio:2020wdo, OHare:2024nmr} for recent reviews.
On the other hand, a significant number of ultra-light ALPs can originate from higher-dimensional gauge fields, including, of course, the QCD axion candidate \cite{Witten:1984dg, Green:1984sg, Choi:2003wr, Reece:2024wrn, Gendler:2024gdo, Li:2024jko}.

Within such a theoretical framework of the axiverse, it is natural to consider the cosmological evolution of multiple axions, a subject that has garnered considerable attention in recent times \cite{Hill:1988bu, Daido:2015cba, Cyncynates:2023esj, Li:2023uvt, Li:2025cep}.
When considering non-zero mass mixing between axions, a significant evolution known as level crossing can occur during the QCD phase transition. 
Typically, level crossing precedes the critical temperature of the QCD phase transition, denoted as $T_{\rm QCD}$. 
However, as emphasized in ref.~\cite{Li:2023uvt}, this rule is not absolute; in scenarios featuring a $Z_{\mathcal N}$ axion,\footnote{In the $Z_{\mathcal N}$ axion scenario \cite{Hook:2018jle, DiLuzio:2021pxd}, ${\mathcal N}$ mirror and degenerate worlds, nonlinearly realized through the axion field under a $Z_{\mathcal N}$ symmetry, can potentially coexist in Nature. Notably, one of these worlds corresponds to the SM world. To address the strong CP problem in this context, it is imperative that ${\mathcal N}$ be an odd integer satisfying ${\mathcal N} \geqslant 3$. The $Z_{\mathcal{N}}$ axion can also serve as a candidate for DM \cite{DiLuzio:2021gos}.} a second level crossing might occur precisely at $T_{\rm QCD}$, a phenomenon termed {\it double level crossings} (see also ref.~\cite{Li:2024kdy} for another illustration). 
Axion level crossing can induce the adiabatic or non-adiabatic transitions in axion energy density, similar to the MSW effect \cite{Wolfenstein:1977ue, Mikheyev:1985zog, Mikheev:1986wj} in neutrino oscillations, and has some profound cosmological implications \cite{Kitajima:2014xla, Daido:2015bva, Ho:2018qur, Cyncynates:2021xzw, Cyncynates:2022wlq, Li:2023xkn, Li:2024psa, Murai:2024nsp, Li:2025uwq}.

In this work, we focus on axion double level crossings within the context of multi-axion mass mixing where the number of axions exceeds {\it two}. 
Firstly, we present the comprehensive model for axion double level crossings, encompassing both the light and heavy axion scenarios. 
Following this, we provide three illustrative toy examples in the light axion scenario. 
Additionally, we redefine the light and heavy axion scenarios within the scope of this work.
We find that double level crossings are a prevalent phenomenon in the mass mixing of the $Z_{\mathcal N}$ axion and multiple ALPs. 
In the light axion scenario, double level crossings can manifest multiple times as ${\mathcal N}$ approaches large values.
However, it is crucial to note that excessively large ${\mathcal N}$ may also prevent the occurrence of double level crossings.
Conversely, in the heavy axion scenario, an excessively small ${\mathcal N}$ may similarly prevent their appearance.

The paper is structured as follows.  
In section~\ref{sec_general_model}, we present the general model for axion double level crossings.
In section~\ref{sec_toy_examples}, we discuss several toy examples in the light axion scenario.
Section~\ref{sec_redefining} redefines the light and heavy axion scenarios.
Section~\ref{sec_implications} briefly discusses the cosmological implications.
Finally, section~\ref{sec_Conclusion} concludes the paper.

\section{The general model for axion double level crossings}
\label{sec_general_model}

In this section, we present the general model for axion double level crossings within the framework of multi-axion mass mixing in the string axiverse \cite{Li:2025cep}.
Considering a scenario where the $Z_{\mathcal N}$ axion ($a_{\mathcal N}$) coexists with multiple ALPs ($A_i$),\footnote{For the original  axion double level crossings involving the mixing scenario of one $Z_{\mathcal N}$ axion and one ALP, please refer to our previous ref.~\cite{Li:2023uvt}.} the low-energy effective Lagrangian for this type of mixing can be formulated as follows
\begin{eqnarray}
\begin{aligned}
\mathcal{L}&\supset\dfrac{1}{2}f_{a_{\mathcal N}}^2\left(\partial\theta_{\mathcal N}\right)^2+\dfrac{1}{2}\sum_{i=1}^{N}f_{A_i}^2\left(\partial\Theta_i\right)^2\\
&-\dfrac{m_{a_{\mathcal N}}^2f_{a_{\mathcal N}}^2}{{\mathcal N}^2}\left[1-\cos\left(n_{00}\theta_{\mathcal N}+\sum_{j=1}^{N}n_{0j}\Theta_j+\delta_0\right)\right]\\
&-\sum_{i=1}^{N}m_{A_i}^2f_{A_i}^2\left[1-\cos\left(n_{i0}\theta_{\mathcal N}+\sum_{j=1}^{N}n_{ij}\Theta_j+\delta_i\right)\right]\, ,
\end{aligned}
\end{eqnarray}
where $\theta_{\mathcal N}$ and $\Theta_i$ are the $Z_{\mathcal N}$ axion and ALP angles, respectively, $f_{a_{\mathcal N}}$ and $f_{A_i}$ the axion decay constants, $m_{a_{\mathcal N}}$ and $m_{A_i}$ the axion masses, $n_{ij}$ the domain wall numbers, and $\delta_i$ the constant phases.
The $Z_{\mathcal N}$ axion mass is detailed in appendix~\ref{appendix_mass}, while the ALP mass remains constant in the simplest single-field scenario.
The constant phases $\delta_i$ are deemed to be zero, which is equivalent to seeking a separate resolution to the strong CP problem.
Since this assumption does not alter the evolution of the axion fields during mixing, it has no impact on the final axion production in our scenario \cite{Li:2023xkn}.

To ensure the occurrence of maximal mass mixing, according to ref.~\cite{Li:2025cep}, we need to assume that the masses of all ALPs $m_{A_i}$ must be less than the intermediate-temperature mass of the $Z_{\mathcal N}$ axion $m_{a_{\mathcal N},\pi}$,\footnote{In the canonical QCD axion case, this is the zero-temperature mass of the QCD axion.} and there must be no equality among the ALP masses, $\rm i.e.$, $m_{A_i}<m_{a_{\mathcal N},\pi},\, \forall i$, and $m_{A_i}\neq m_{A_j},\, \forall i\neq j$. 
This is because, in the case of equality, effective mixing would occur only once. 
Furthermore, the decay constants of all ALPs $f_{A_i}$ must be either less than or greater than the decay constant of the $Z_{\mathcal N}$ axion $f_{a_{\mathcal N}}$ simultaneously, $\rm i.e.$, $f_{A_i}<f_{a_{\mathcal N}},\, \forall i$ or $f_{A_i}>f_{a_{\mathcal N}},\, \forall i$.
Depending on different selections of the axion decay constants, the mixing discussed here can typically be further categorized into the {\it light axion scenario} \cite{Daido:2015cba, Li:2023uvt}, which suppresses the (QCD or $Z_{\mathcal N}$) axion energy density, and the {\it heavy axion scenario} \cite{Cyncynates:2023esj}, which enhances the axion energy density.\footnote{We will redefine the light and heavy axion scenarios in this work. In fact, they should be categorized as $f_{A_i}<f_{a_{\mathcal N}}/{\mathcal N},\, \forall i$ and $f_{A_i}>f_{a_{\mathcal N}}/{\mathcal N},\, \forall i$, respectively.} 
Here, we first discuss the light axion scenario, where the decay constants of all ALPs are smaller than that of the $Z_{\mathcal N}$ axion.
In this case, the matrix of domain wall numbers $n_{ij}$ should be taken as follows
\begin{eqnarray}
\mathfrak{n}_{\mathfrak{ij}}=
\left(
\begin{array}{cccccc}
{\mathcal N}  & ~ 0 & ~ 0 & ~ 0 &~\cdots & ~ 0\\
{\mathcal N}  & ~ 1 & ~ 0 & ~ 0 &~\cdots & ~ 0\\
{\mathcal N}  & ~ 0 & ~ 1 & ~ 0 &~\cdots & ~ 0\\
{\mathcal N}  & ~ 0 & ~ 0 & ~ 1 &~\cdots & ~ 0\\
\vdots  & ~\vdots & ~\vdots &~\vdots & ~\ddots & ~ \vdots\\
{\mathcal N}  & ~ 0 & ~ 0 & ~ 0 & ~ \cdots & ~ 1
\end{array}
\right)\, .
\end{eqnarray}
Notice that $\mathfrak{n}_{\mathfrak{ij}}$ is a $(N+1)\times(N+1)$ matrix, beginning with indices $\mathfrak{i}=0$ and $\mathfrak{j}=0$.
The mass mixing matrix is given by
\begin{eqnarray}
\mathbf{M}^2=
\left(
\begin{array}{ccccc}
m_{a_{\mathcal N}}^2+\dfrac{{\mathcal N}^2}{f_{a_{\mathcal N}}^2} \displaystyle\sum_{i=1}^{N} m_{A_i}^2 f_{A_i}^2& ~ \dfrac{{\mathcal N}m_{A_1}^2 f_{A_1}}{f_{a_{\mathcal N}}} & ~ \dfrac{{\mathcal N}m_{A_2}^2 f_{A_2}}{f_{a_{\mathcal N}}} & ~\cdots &~ \dfrac{{\mathcal N}m_{A_N}^2 f_{A_N}}{f_{a_{\mathcal N}}}\\
\dfrac{{\mathcal N}m_{A_1}^2 f_{A_1}}{f_{a_{\mathcal N}}}   & ~ m_{A_1}^2 & ~ 0 & ~\cdots & ~0\\
\dfrac{{\mathcal N}m_{A_2}^2 f_{A_2}}{f_{a_{\mathcal N}}}   & ~ 0 & ~ m_{A_2}^2 & ~\cdots & ~0\\
\vdots  & ~\vdots & ~\vdots  & ~\ddots & ~\vdots\\
\dfrac{{\mathcal N}m_{A_N}^2 f_{A_N}}{f_{a_{\mathcal N}}}   & ~ 0 & ~ 0 & ~\cdots & ~ m_{A_N}^2
\end{array}
\right)\, .
\end{eqnarray}
Since we assume that the masses of the ALPs are all unequal, the mass mixing matrix here can be expressed in the form of multiple effective matrices \cite{Li:2025cep}. 
Specifically, it can be represented as
\begin{eqnarray}
\mathbf{M}_i^2=
\left(\begin{array}{cc}
m_{a_{\mathcal N}}^2+\dfrac{{\mathcal N}^2 m_{A_i}^2 f_{A_i}^2}{f_{a_{\mathcal N}}^2}  & ~ \dfrac{{\mathcal N}m_{A_i}^2 f_{A_i}}{f_{a_{\mathcal N}}}\\
\dfrac{{\mathcal N}m_{A_i}^2 f_{A_i}}{f_{a_{\mathcal N}}} & ~ m_{A_i}^2
\end{array}\right)\, ,
\end{eqnarray} 
with the corresponding mass eigenvalues $m_{h_i,l_i}$. 
Then we can obtain the effective mass eigenvalues $m_{e_i}$ and the first level crossing temperatures $T_{\times_i}$, as shown in appendix~\ref{appendix_temperatures}.
While in the context of the heavy axion scenario, where the decay constants of all ALPs are greater than that of the $Z_{\mathcal N}$ axion, the matrix of domain wall numbers should be considered as follows
\begin{eqnarray}
\mathfrak{n}_{\mathfrak{ij}}=
\left(
\begin{array}{cccccc}
{\mathcal N}  & ~ 1 & ~ 1 & ~ 1 &~\cdots & ~ 1\\
0  & ~ 1 & ~ 0 & ~ 0 &~\cdots & ~ 0\\
0  & ~ 0 & ~ 1 & ~ 0 &~\cdots & ~ 0\\
0  & ~ 0 & ~ 0 & ~ 1 &~\cdots & ~ 0\\
\vdots  & ~\vdots & ~\vdots &~\vdots & ~\ddots & ~ \vdots\\
0  & ~ 0 & ~ 0 & ~ 0 & ~ \cdots & ~ 1
\end{array}
\right)\, .
\end{eqnarray}
In this case, the mass mixing matrix is given by
\begin{eqnarray}
\mathbf{M}^2=
\left(
\begin{array}{ccccc}
m_{a_{\mathcal N}}^2 & ~ \dfrac{m_{a_{\mathcal N}}^2 f_{a_{\mathcal N}}}{{\mathcal N}f_{A_1}^{}} & ~ \dfrac{m_{a_{\mathcal N}}^2 f_{a_{\mathcal N}}}{{\mathcal N}f_{A_2}} & ~\cdots &~ \dfrac{m_{a_{\mathcal N}}^2 f_{a_{\mathcal N}}}{{\mathcal N}f_{A_N}}\\
\dfrac{m_{a_{\mathcal N}}^2 f_{a_{\mathcal N}}}{{\mathcal N}f_{A_1}}   & ~ \dfrac{m_{a_{\mathcal N}}^2 f_{a_{\mathcal N}}^2}{{\mathcal N}^2f_{A_1}^2}+m_{A_1}^2 & ~ \dfrac{m_{a_{\mathcal N}}^2 f_{a_{\mathcal N}}^2}{{\mathcal N}^2f_{A_1} f_{A_2}} & ~\cdots & ~\dfrac{m_{a_{\mathcal N}}^2 f_{a_{\mathcal N}}^2}{{\mathcal N}^2f_{A_1} f_{A_N}}\\
\dfrac{m_{a_{\mathcal N}}^2 f_{a_{\mathcal N}}}{{\mathcal N}f_{A_2}}   & ~ \dfrac{m_{a_{\mathcal N}}^2 f_{a_{\mathcal N}}^2}{{\mathcal N}^2f_{A_1} f_{A_2}} & ~ \dfrac{m_{a_{\mathcal N}}^2 f_{a_{\mathcal N}}^2}{{\mathcal N}^2f_{A_2}^2}+m_{A_2}^2 & ~\cdots & ~\dfrac{m_{a_{\mathcal N}}^2 f_{a_{\mathcal N}}^2}{{\mathcal N}^2f_{A_2} f_{A_N}}\\
\vdots  & ~\vdots & ~\vdots  & ~\ddots & ~ \vdots\\
\dfrac{m_{a_{\mathcal N}}^2 f_{a_{\mathcal N}}}{{\mathcal N}f_{A_N}}   & ~ \dfrac{m_{a_{\mathcal N}}^2 f_{a_{\mathcal N}}^2}{{\mathcal N}^2f_{A_1} f_{A_N}} & ~ \dfrac{m_{a_{\mathcal N}}^2 f_{a_{\mathcal N}}^2}{{\mathcal N}^2f_{A_2} f_{A_N}} & ~ \cdots & ~ \dfrac{m_{a_{\mathcal N}}^2 f_{a_{\mathcal N}}^2}{{\mathcal N}^2f_{A_N}^2}+m_{A_N}^2
\end{array}
\right)\, ,
\end{eqnarray} 
with the the effective mixing matrices
\begin{eqnarray}
\mathbf{M}_i^2=
\left(\begin{array}{cc}
m_{a_{\mathcal N}}^2  & ~ \dfrac{m_{a_{\mathcal N}}^2 f_{a_{\mathcal N}}}{{\mathcal N}f_{A_i}}\\
\dfrac{m_{a_{\mathcal N}}^2 f_{a_{\mathcal N}}}{{\mathcal N}f_{A_i}} & ~ \dfrac{m_{a_{\mathcal N}}^2 f_{a_{\mathcal N}}^2}{{\mathcal N}^2f_{A_i}^2}+m_{A_i}^2
\end{array}\right)\, ,
\end{eqnarray}
and the corresponding mass eigenvalues $m_{h_i,l_i}$.
See also appendix~\ref{appendix_temperatures} for the first level crossing temperatures $T_{\times_i}$ in this scenario. 

Here, we have considered a series of models for axion double level crossings, where the number of ALPs ($n_{\rm ALP}$) is no less than one. 
The most basic scenario, with $n_{\rm ALP}=1$, was discussed in ref.~\cite{Li:2023uvt}, where the concept of double level crossings was first proposed. 
Notice that the ``double$"$ level crossings discussed here should not be confused with the ``multiple$"$ level crossings referred to in ref.~\cite{Li:2025cep}, which actually involves multiple instances of single level crossing, whereas here it specifically refers to scenarios with multiple instances of double level crossings. 
In the next section, we will provide several illustrative examples of the case where $n_{\rm ALP}=2$.

%%%%%%%%%%%%%%%%%%%%%%%%%%
\begin{figure}[h] 
\centering
\includegraphics[width=0.5915\textwidth]{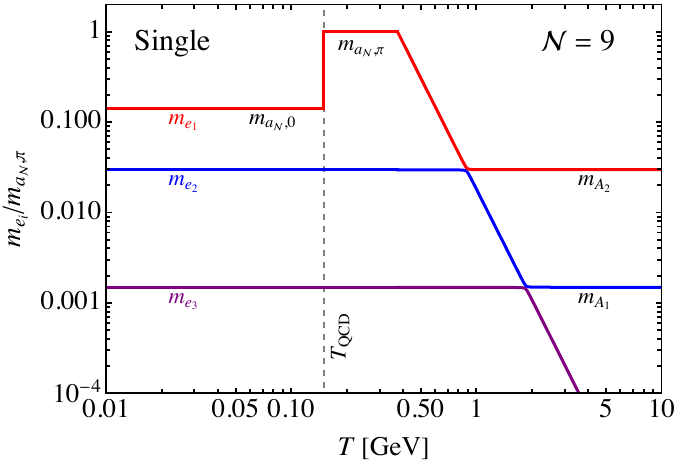}
\includegraphics[width=0.5915\textwidth]{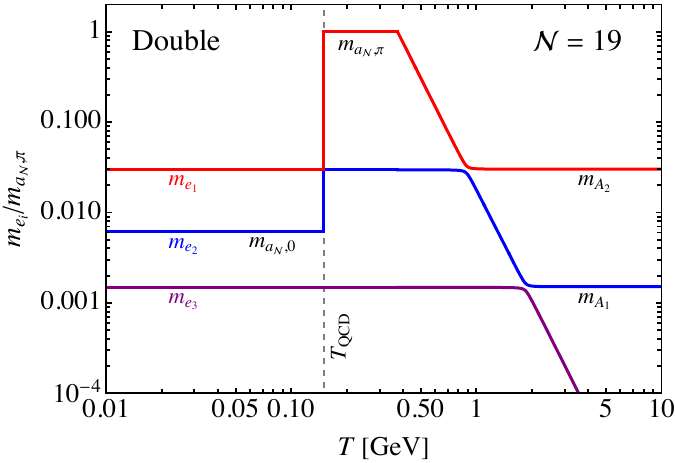}
\includegraphics[width=0.5915\textwidth]{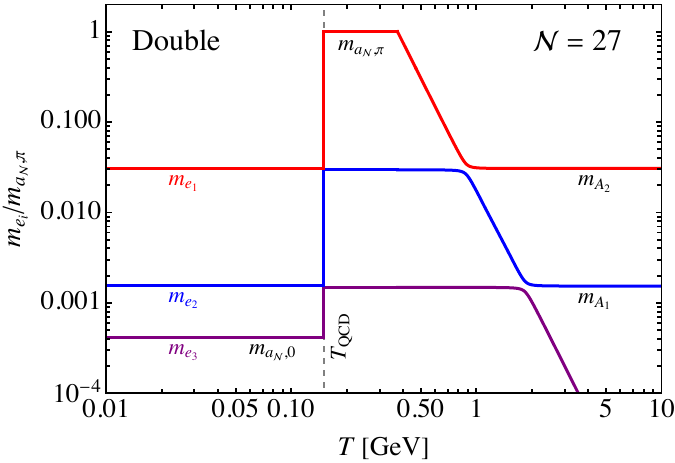}
\caption{Toy examples of axion double level crossings with $n_{\rm ALP}=2$.
The model parameter settings can be found in table~\ref{tab_1}.
See also the main text for more details.}
\label{fig_me}
\end{figure}   
  
%%%%%%%%%%%%%%%%%%%%%%%%%%
\begin{table}[t]
\centering
\begin{tabular}{lcccccc}
\hline\hline
Plots   &    $f_{a_{\mathcal N}}$ ($\rm GeV$)   &    $f_{A_{1,2}}$ ($\rm GeV$)  &   $m_{A_1}/m_{a_{\mathcal N},\pi}$ & $m_{A_2}/m_{a_{\mathcal N},\pi}$ & $\gamma$ & ${\mathcal N}$\\
\hline
top       &  $1\times10^{10}$     &  $1\times10^{8}$   &  $1.5\times10^{-3}$  &  $3\times10^{-2}$  & 0.4 & 9 \\
middle &  $1\times10^{10}$     &  $1\times10^{8}$   &  $1.5\times10^{-3}$  &  $3\times10^{-2}$   & 0.4 & 19 \\
bottom &  $1\times10^{10}$     &  $1\times10^{8}$   &  $1.5\times10^{-3}$  &  $3\times10^{-2}$   & 0.4 & 27 \\
 \hline\hline
\end{tabular}
\caption{The model parameter settings for the toy examples presented in figure~\ref{fig_me}.
Notice that we set $f_{A_1}=f_{A_2}$.}
\label{tab_1}
\end{table}   
  
\section{Toy examples of axion double level crossings with $n_{\rm ALP}=2$} 
\label{sec_toy_examples}
 
In this section, we present several toy examples of axion double level crossings in the light axion scenario with $n_{\rm ALP}=2$, see figure~\ref{fig_me}.
The solid lines represent the normalized temperature-dependent axion mass eigenvalues $m_{e_i}/m_{a_{\mathcal N},\pi}$ as functions of the cosmic temperature $T$. 
The model parameter settings can be found in table~\ref{tab_1}.
Notice that we set $m_{A_1}<m_{A_2}$ and $f_{A_i}\ll f_{a_{\mathcal N}}$ ($\rm i.e.$, the light axion scenario).
For a clearer comparison, here we only need to vary the value of ${\mathcal N}$ to achieve the purpose of illustration. 
The top, middle, and bottom plots of this figure respectively represent three typical types of level crossing phenomena:
\begin{itemize}
\item {\bf Single level crossing.} 
The top plot shows a small ${\mathcal N}$ scenario where the zero-temperature $Z_{\mathcal N}$ axion mass $m_{a_{\mathcal N},0}$ is larger than the ALP mass $m_{A_2}$,
\begin{eqnarray}
m_{A_2}<\dfrac{m_\pi f_\pi}{\sqrt[4]{\pi} f_{a_{\mathcal N}}}\sqrt[4]{\dfrac{1-z}{1+z}}{\mathcal N}^{3/4}z^{\mathcal N/2}\, ,
\end{eqnarray}
where $m_\pi$ and $f_\pi$ represent the mass and decay constant of the pion, respectively, and $z\equiv m_u/m_d\simeq0.48$ represents the ratio of the up to down quark masses.
The essence of this scenario remains a single level crossing, even though the level crossing occurs twice at the corresponding level crossing temperatures $T_{\times_1}$ and $T_{\times_2}$, respectively.
\item {\bf Single \& double level crossings.} 
In the middle plot, we show a scenario where the zero-temperature $Z_{\mathcal N}$ axion mass $m_{a_{\mathcal N},0}$ is smaller than the ALP mass $m_{A_2}$, but larger than another ALP mass $m_{A_1}$, 
\begin{eqnarray}
m_{A_1}<\dfrac{m_\pi f_\pi}{\sqrt[4]{\pi} f_{a_{\mathcal N}}}\sqrt[4]{\dfrac{1-z}{1+z}}{\mathcal N}^{3/4}z^{\mathcal N/2}<m_{A_2}\, .
\end{eqnarray} 
Here, we observe that the double level crossings phenomenon occurs between the mass eigenvalues $m_{e_1}$ and $m_{e_2}$, with the first happening at the level crossing temperature $T_{\times_2}$ and the second at the QCD phase transition critical temperature $T_{\rm QCD}$.
Moreover, there exists a single level crossing between the mass eigenvalues $m_{e_2}$ and $m_{e_3}$, which occurs at the level crossing temperature $T_{\times_1}$.
\item {\bf Multiple double level crossings.}
In addition, the bottom plot shows a large ${\mathcal N}$ scenario where the zero-temperature $Z_{\mathcal N}$ axion mass $m_{a_{\mathcal N},0}$ is smaller than the ALP mass $m_{A_1}$,
\begin{eqnarray}
m_{A_1}>\dfrac{m_\pi f_\pi}{\sqrt[4]{\pi} f_{a_{\mathcal N}}}\sqrt[4]{\dfrac{1-z}{1+z}}{\mathcal N}^{3/4}z^{\mathcal N/2}\, .
\end{eqnarray} 
Based on the middle plot above, we observe that the double level crossings phenomenon also occurs between the mass eigenvalues $m_{e_2}$ and $m_{e_3}$, and similarly, the second level crossing occurs at $T_{\rm QCD}$.
\end{itemize} 
These three plots effectively illustrate three distinct types of level crossing scenarios, among which the last two, representing {\it multiple} double level crossings, are presented for the first time.
With other model parameters remaining constant, this can be achieved by altering the value of ${\mathcal N}$, which corresponds to the relationship between the zero-temperature $Z_{\mathcal N}$ axion mass $m_{a_{\mathcal N},0}$ and the ALP masses $m_{A_i}$.
We observe that, in principle, the larger the value of ${\mathcal N}$, the more frequent the occurrence of the double level crossings. 
However, this is not always the case, as the conditions for level crossing are influenced by ${\mathcal N}$. 
From figure~\ref{fig_me}, we can see that as ${\mathcal N}$ increases, the first level crossing may be compromised. 
In the next section, we will delve into this issue.
Notice that what is presented here pertains specifically to the light axion scenario. 
Although the evolution of axions in the heavy axion scenario appears similar, there will be notable differences in the conditions under which the level crossing occurs and in the transition of axion energy density.

%%%%%%%%%%%%%%%%%%%%%%%%%%
\begin{figure}[t]
\centering
\includegraphics[width=0.70\textwidth]{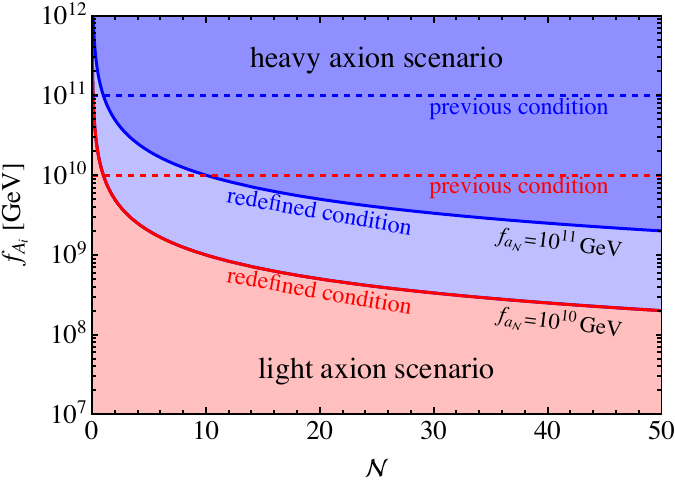}
\caption{Redefined conditions for the light and heavy axion scenarios.
The solid red and blue lines represent $f_{a_{\mathcal N}}=1\times10^{10}\, \rm GeV$ and $1\times10^{11}\, \rm GeV$, respectively. 
The area below the solid line represents the redefined light axion scenario, whereas the area above it represents the redefined heavy axion scenario. 
The dashed red and blue lines correspond to the previous conditions.}
\label{fig_redefined_conditions}
\end{figure}  

%%%%%%%%%%%%%%%%%%%%%%%%%%
\begin{figure}[t]
\centering
\includegraphics[width=0.7\textwidth]{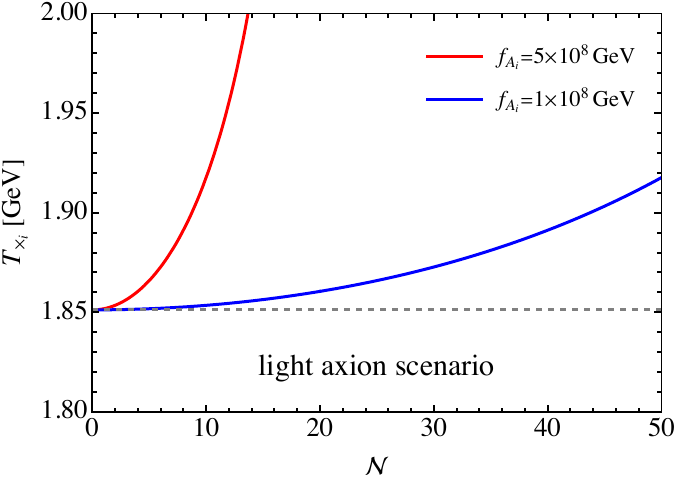}
\includegraphics[width=0.7\textwidth]{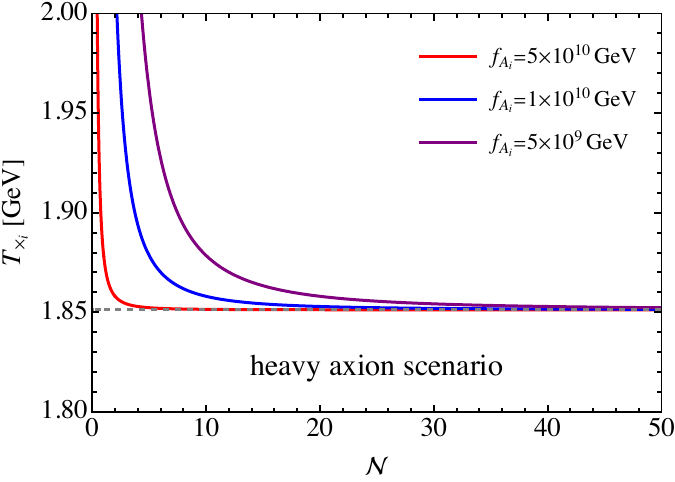}
\caption{The first level crossing temperatures $T_{\times_i}$ in the light (top) and heavy (bottom) axion scenarios.
Here we set $f_{a_{\mathcal N}}=1\times10^{10}\, \rm GeV$, $m_{A_i}/m_{a_{\mathcal N},\pi}=1.5\times10^{-3}$, and $\gamma=0.4$.
The gray dashed lines represent the extreme scenarios where the values of $f_{A_i}$ and $f_{a_{\mathcal N}}$ differ greatly.}
\label{fig_Txi}
\end{figure}

\section{Redefining the light and heavy axion scenarios} 
\label{sec_redefining}
 
Above, we demonstrate the phenomenon of axion double level crossings in the light axion scenario where $n_{\rm ALP}=2$.
Here we discuss some issues in the more general case, where $n_{\rm ALP}>2$. 
In the context of light and heavy axion scenarios, the temperatures of the first level crossing can be expressed as (see also appendix~\ref{appendix_temperatures})
\begin{eqnarray}
\text{Light:}&&
T_{\times_i}=\dfrac{T_{\rm QCD}}{\gamma} \left(\dfrac{\left(1-z^2\right)\left(f_{a_{\mathcal N}}^2-{\mathcal N}^2 f_{A_i}^2\right)m_{A_i}^2}{z m_\pi^2 f_\pi^2}\right)^{-\frac{1}{2b}}\, ,\\
\text{Heavy:}&&
T_{\times_i}=\dfrac{T_{\rm QCD}}{\gamma} \left(\dfrac{\left(1-z^2\right)\left({\mathcal N}^2 f_{A_i}^2-f_{a_{\mathcal N}}^2\right){\mathcal N}^2 m_{A_i}^2 f_{A_i}^2 f_{a_{\mathcal N}}^2}{z m_\pi^2 f_\pi^2 \left({\mathcal N}^2 f_{A_i}^2+f_{a_{\mathcal N}}^2\right)^2}\right)^{-\frac{1}{2b}}\, ,
\end{eqnarray}
respectively. 
As mentioned earlier, the value of ${\mathcal N}$ in the light axion scenario cannot be infinitely large, as it would violate the conditions for the first level crossing to occur. 
This can be obtained from the condition $f_{a_{\mathcal N}}^2-{\mathcal N}^2 f_{A_i}^2>0$, and thus we can redefine the light axion scenario with\footnote{Although this similar condition was mentioned in ref.~\cite{Li:2023uvt} in the context of two-axion mixing, the subsequent heavy axion scenario was not discussed in that paper.}
\begin{eqnarray} 
f_{A_i}<\dfrac{f_{a_{\mathcal N}}}{{\mathcal N}},\, \forall i\, ,
\end{eqnarray}
which is a stronger condition than $f_{A_i}<f_{a_{\mathcal N}}$, especially in the large ${\mathcal N}$ limit.
Similarly, we can also redefine the heavy axion scenario based on the condition ${\mathcal N}^2 f_{A_i}^2-f_{a_{\mathcal N}}^2>0$, and then we have  
\begin{eqnarray} 
f_{A_i}>\dfrac{f_{a_{\mathcal N}}}{{\mathcal N}},\, \forall i\, .
\end{eqnarray}
We find that this is a weaker condition than $f_{A_i}>f_{a_{\mathcal N}}$, allowing $f_{A_i}\lesssim f_{a_{\mathcal N}}$, as long as the value of ${\mathcal N}$ is not too small.
In figure~\ref{fig_redefined_conditions}, we show the redefined conditions for the light and heavy axion scenarios in the $\{{\mathcal N},\, f_{A_i}\}$ plane.
The area below the solid line represents the newly defined light axion scenario, whereas the area above it represents the newly defined heavy axion scenario. 
On the other hand, the dashed red and blue lines correspond to the previous conditions.
It is intuitive from this figure that the redefined light axion scenario is constrained by the large values of ${\mathcal N}$. 
In the redefined heavy axion scenario, an interesting aspect is that $f_{A_i}$ can be equal to or even smaller than $f_{a_{\mathcal N}}$, which may have some intriguing implications, particularly regarding the axion energy density. 
This difference arises because, in previous considerations of axion level crossing, $f_{A_i}$ are typically either less than or greater than $f_{a_{\mathcal N}}$.

Finally, we also show in figure~\ref{fig_Txi} the first level crossing temperatures in both the light and heavy axion scenarios.
The solid lines represent the temperatures $T_{\times_i}$ as functions of the value of ${\mathcal N}$. 
The gray dashed lines represent the extreme scenarios where the values of $f_{A_i}$ and $f_{a_{\mathcal N}}$ differ greatly, $\rm i.e.$, $f_{A_i}\ll f_{a_{\mathcal N}}$ and $f_{A_i}\gg f_{a_{\mathcal N}}$, respectively.
We find that, in the light axion scenario, variations in $T_{\times_i}$ exhibit sensitivity to large values of ${\mathcal N}$, whereas, in contrast, in the heavy axion scenario, they are sensitive to small values of ${\mathcal N}$.
Notice that here we have selected some typical values of $f_{A_i}$ for illustration, particularly in the heavy axion scenario, where we have chosen three sets of $f_{A_i}$ such that they can be greater than, equal to, and less than $f_{a_{\mathcal N}}$, respectively.
Therefore, in the light axion scenario, a larger value ${\mathcal N}$ corresponds to a higher first level crossing temperature. 
When this temperature deviates significantly from that indicated by the gray dashed line, we consider the first level crossing to be compromised, thereby affecting the realization of double level crossings. 
Conversely, in the heavy axion scenario, the first level crossing can be compromised for smaller values of ${\mathcal N}$, which similarly hinders the realization of double level crossings.

\section{A brief discussion on cosmological implications} 
\label{sec_implications}

In this section, we briefly discuss the cosmological implications of axion double level crossings presented in this work.

Axion level crossing carries profound cosmological implications, encompassing axion relic density, isocurvature fluctuations, dark energy, domain walls, gravitational waves, and primordial black holes, as detailed in ref.~\cite{Li:2025cep}. 
Similarly, the axion double level crossings investigated here entail analogous cosmological consequences. 
While a comprehensive exploration of these effects is beyond the scope of this work and remains a promising avenue for future research, we briefly highlight below the novel insights our study offers regarding these phenomena.

Of these cosmological consequences, the modification of the axion relic density stands out as the most direct and crucial, serving as the foundation upon which the other effects depend. 
In this work, the parameter ${\mathcal N}$ plays a pivotal role. 
On the one hand, as previously discussed, the light and heavy axion scenarios impose upper and lower bounds on ${\mathcal N}$, respectively. 
On the other hand, the $Z_{\mathcal N}$ axion model and its corresponding DM solution inherently constrain ${\mathcal N}$, a restriction manifested in the $\{m_{a_{\mathcal N}}, 1/f_{a_{\mathcal N}}\}$ parameter plane. 
Consequently, we can constrain ${\mathcal N}$ by evaluating the degree of deviation from the ideal first level crossing, thereby narrowing the viable parameter space for the axion mass $m_{a_{\mathcal N}}$ and decay constant $f_{a_{\mathcal N}}$.
Furthermore, here we consider the mixing of multiple ALPs with the $Z_{\mathcal N}$ axion. 
Given that ALP masses must be non-degenerate and exhibit a hierarchical structure, the number of ALPs participating in the double level crossings provides an additional constraint on ${\mathcal N}$. 
Conversely, this limits the ALP parameter space, particularly the decay constants $f_{A_i}$ of ultra-light ALPs in the heavy axion scenario, where $f_{A_i}>f_{a_{\mathcal N}}/{\mathcal N}$. 
Notably, these ALPs fall within the sensitivity reach of future axion detection experiments, such as CASPEr-electric \cite{Budker:2013hfa} and proposals based on the piezoaxionic effect \cite{Arvanitaki:2021wjk}.
Numerous additional searches for ALP couplings --- specifically to photons, electrons, and nucleons --- are also ongoing, though they are not discussed further here.
Finally, while it is possible to directly calculate the relic density of the mixed $Z_{\mathcal N}$ axion or ALPs via the misalignment mechanism for a given model, such calculations are of secondary importance in this context due to the large number of model parameters, many of which remain undetermined.

\section{Conclusion}%%%%%%%%%%%%%%%%%%%%%%%%%%%Conclusion 
\label{sec_Conclusion}

In summary, we have investigated axion double level crossings within the context of multi-axion mass mixing where the number of axions exceeds two. 
We find that double level crossings are a prevalent phenomenon in the mass mixing of the $Z_{\mathcal N}$ axion and multiple ALPs. 
Firstly, we present the comprehensive model for axion double level crossings, encompassing both the light and heavy axion scenarios. 
The corresponding matrix of domain wall numbers and mass mixing matrix are shown in detail.
Subsequently, we provide three toy examples in the light axion scenario, which effectively illustrate three distinct types of level crossing scenarios.
Additionally, we redefine the light and heavy axion scenarios within the scope of this work, categorizing them as $f_{A_i}<f_{a_{\mathcal N}}/{\mathcal N},\, \forall i$ and $f_{A_i}>f_{a_{\mathcal N}}/{\mathcal N},\, \forall i$, respectively.
In the light axion scenario, a large ${\mathcal N}$ can cause multiple double level crossings, but an excessively large ${\mathcal N}$ may prevent them. 
While in the heavy axion scenario, too small ${\mathcal N}$ can similarly prevent double level crossings.
Our study of axion double level crossings has provided a deeper understanding of axion physics and its potential implications for cosmology. 
 
\section*{Acknowledgments}%%%%%%%%%%%%%%%%Acknowledgments

H.J.L. would like to express his gratitude to the International Centre for Theoretical Physics Asia-Pacific for its hospitality, as part of this work was completed there.
W.C. is supported by the National Key R\&D Program of China (Grant No.~2023YFA1607104), the National Natural Science Foundation of China (NSFC) (Grants No.~11775025 and No.~12175027), and the Fundamental Research Funds for the Central Universities (Grant No.~2017NT17).
H.K.G. is supported by the NSFC (Grant No.~12347103).
Y.F.Z. is supported by the CAS Project for Young Scientists in Basic Research YSBR-006, the National Key R\&D Program of China (Grant No.~2017YFA0402204), and the NSFC (Grants No.~11821505, No.~11825506, and No.~12047503).

\appendix

\section{The $Z_{\mathcal N}$ axion mass}
\label{appendix_mass}

The temperature-dependent $Z_{\mathcal N}$ axion mass $m_{a_{\mathcal N}}(T)$ can be described by \cite{DiLuzio:2021pxd, DiLuzio:2021gos}
\begin{eqnarray}
m_{a_{\mathcal N}}(T)\simeq
\begin{cases}
\dfrac{m_\pi f_\pi}{\sqrt[4]{\pi} f_{a_{\mathcal N}}}\sqrt[4]{\dfrac{1-z}{1+z}}{\mathcal N}^{3/4}z^{\mathcal N/2}\, , &\text{for } T\leq T_{\rm QCD}\\
\dfrac{m_\pi f_\pi}{f_{a_{\mathcal N}}}\sqrt{\dfrac{z}{1-z^2}}\, , &\text{for } T_{\rm QCD} < T \leq \dfrac{T_{\rm QCD}}{\gamma}\\
\dfrac{m_\pi f_\pi}{f_{a_{\mathcal N}}}\sqrt{\dfrac{z}{1-z^2}}\left(\dfrac{\gamma T}{T_{\rm QCD}}\right)^{-b}\, , &\text{for } T>\dfrac{T_{\rm QCD}}{\gamma}
\end{cases} 
\label{m_ZNaxion}
\end{eqnarray}
where $m_\pi$ and $f_\pi$ represent the mass and decay constant of the pion, respectively, $z\equiv m_u/m_d\simeq0.48$ represents the ratio of the up ($m_u$) to down ($m_d$) quark masses, $T_{\rm QCD}\simeq 150\, \rm MeV$ represents the critical temperature of the QCD phase transition, and $\gamma\in(0,1)$ is a temperature parameter, which we set to $\gamma=0.4$ in this work.
The first and second terms in eq.~\eqref{m_ZNaxion} correspond to the zero-temperature $Z_{\mathcal N}$ axion mass $m_{a_{\mathcal N},0}$ and the intermediate-temperature mass $m_{a_{\mathcal N},\pi}$, respectively.
See also ref.~\cite{DiLuzio:2021gos} for more details on $Z_{\mathcal N}$ axion mass.

\section{The first level crossing temperatures}
\label{appendix_temperatures}

Here we present the mass eigenvalues in the context of multi-axion mass mixing \cite{Li:2025cep}, and also show the corresponding first level crossing temperatures obtained in this work.
If we examine a scenario in which the ALP mass of $A_i$ is less than that of $A_{i+1}$, we have the mass sequence
\begin{eqnarray}
m_{A_i}<m_{a_{\mathcal N},\pi},\, \forall i\, ,\\
m_{A_1}<m_{A_2}<\cdots<m_{A_N}\, .
\end{eqnarray}
It should be noted that, within the context under consideration, the masses of ALPs are assumed to remain constant. 
This configuration represents the most straightforward single-field ALP model.
Then the mass eigenvalues $m_{e_i}$ in this scenario can be formulated as follows
\begin{eqnarray}
m_{e_1}&=&m_{h_N}\, ,\\
m_{e_2}&\simeq&
\begin{cases}
m_{l_N}\,, &\text{if } T \le T_{\times_{N-1}}^{(m)}\\
m_{h_{N-1}}\,, &\text{if } T > T_{\times_{N-1}}^{(m)}
\end{cases}\\
m_{e_3}&\simeq&
\begin{cases}
m_{l_{N-1}}\, ,& \text{if } T \le T_{\times_{N-2}}^{(m)}\\
m_{h_{N-2}}\, ,& \text{if } T > T_{\times_{N-2}}^{(m)}
\end{cases}\\
&\vdots\nonumber\\
m_{e_N}&\simeq&
\begin{cases}
m_{l_2}\, , &\text{if } T \le T_{\times_1}^{(m)}\\
m_{h_1}\, , &\text{if } T > T_{\times_1}^{(m)}
\end{cases}\\
m_{e_{N+1}}&=&m_{l_1}\, ,
\end{eqnarray}
with the condition $T_{\times_{i+1}}<T_{\times_i}^{(m)}<T_{\times_i}$, where $T_{\times_i}^{(m)}$ represents an intermediate temperature between the first level crossing temperatures $T_{\times_i}$.
Notice that this applies to both light and heavy axion scenarios.
In this case, one can determine the level crossing temperatures by solving the following differential equations
\begin{eqnarray}
\dfrac{d}{dT}\left(m_{e_N}^2(T)-m_{e_{N+1}}^2(T)\right)\Big|_{T>T_{\times_1}^{(m)}}&=&0\, , \quad \text{for }i=1\\
\dfrac{d}{dT}\left(m_{e_{N-i+1}}^2(T)-m_{e_{N-i+2}}^2(T)\right)\Big|_{T_{\times_i}^{(m)}<T<T_{\times_{i-1}}^{(m)}}&=&0\, , \quad \text{for }1< i< N\\
\dfrac{d}{dT}\left(m_{e_1}^2(T)-m_{e_2}^2(T)\right)\Big|_{T_{\rm QCD}<T<T_{\times_{N-1}}^{(m)}}&=&0\, . \, \quad \text{for }i=N
\end{eqnarray}
Subsequently, we can obtain the corresponding first level crossing temperatures in the light axion scenario 
\begin{eqnarray}
T_{\times_i}=\dfrac{T_{\rm QCD}}{\gamma} \left(\dfrac{\left(1-z^2\right)\left(f_{a_{\mathcal N}}^2-{\mathcal N}^2 f_{A_i}^2\right)m_{A_i}^2}{z m_\pi^2 f_\pi^2}\right)^{-\frac{1}{2b}}\, ,
\end{eqnarray}
and in the heavy axion scenario
\begin{eqnarray}
T_{\times_i}=\dfrac{T_{\rm QCD}}{\gamma} \left(\dfrac{\left(1-z^2\right)\left({\mathcal N}^2 f_{A_i}^2-f_{a_{\mathcal N}}^2\right){\mathcal N}^2 m_{A_i}^2 f_{A_i}^2 f_{a_{\mathcal N}}^2}{z m_\pi^2 f_\pi^2 \left({\mathcal N}^2 f_{A_i}^2+f_{a_{\mathcal N}}^2\right)^2}\right)^{-\frac{1}{2b}}\, .
\end{eqnarray} 

\bibliographystyle{JHEP}
\bibliography{references}

\end{document}